\documentclass[aps,preprint]{revtex4}%
\usepackage{amsfonts}
\usepackage{amsmath}
\usepackage{amssymb}
\usepackage{graphicx}%
\providecommand{\U}[1]{\protect\rule{.1in}{.1in}}

\begin{document}
\title{{\LARGE GENERAL RELATIVISTIC TREATMENT OF THE PIONEERS ANOMALY}\\Marcelo Samuel Berman \ and Fernando de Mello Gomide\\Instituto Albert Einstein/Latinamerica\\Av. Candido Hartmann 575 \#17\\80730-440 Curitiba PR Brazil\\msberman@institutoalberteinstein.org}
\begin{abstract}
We consider a General Relativistic generalized RW%
\'{}%
s metric,and find a field of Universal \ rotational global \ centripetal
acceleration, numerically coincident with the value of the Pioneers Anomalous
one.Related subjects are also treated.The rotation defined here is different
from older frameworks, because we propose a Gaussian metric, whose tri-space
rotates relative to the time orthogonal axis, globally.


Dated (last version) 25July2011

PACS:

\end{abstract}
\maketitle

\begin{center}
{\LARGE GENERAL RELATIVISTIC TREATMENT OF THE PIONEERS ANOMALY}

Marcelo Samuel Berman

\bigskip

\end{center}

\bigskip1.INTRODUCTION

\bigskip Attempts to ascribe a rotational state to the Universe, were
carefully described by Godlowski (2011). However, he confessed that there was
no theoretical framework, within General Relativity, to guide the
observations.In the present paper,such a mechanism is provided.The metric to
be presented, makes the tri-dimensional space, globally rotate relative to the
orthogonal time axis.We are now proposing a novel idea, a generalized Gaussian
metric, which is minimally different from the Robertson-Walker%
\'{}%
s one.In Berman (2007), a semi-relativistic treatment, based on the zero-total
energy of the (rotating) Universe, made us conclude that the Pioneers
anomalous deceleration, was a kind of peculiar centripetal effect of the
rotation of the Universe, that could be observed by any cosmological
observer.In the present paper, we prove the alleged zero-total energy of the
rotating  Universe, and supply the metric for such rotation with expansion.We
keep a perfect fluid model,unlike Raychaudhuri%
\'{}%
s vorticities, and we also differ from the metrical rotational states, derived
from non-diagonalized metrics.We shall find an energy-density solution, very
similar to the Berman (2007) solution.As Berman and Gomide (2011a) have shown,
by our framework, \ of a rotating Universe, we explain the three NASA
anomalies, namely, the Pioneers linear deceleration, the spin-down of the
spacecraft when they were undisturbed, and the fly-by.The present paper,
yields a Machian \ solution, while the \ other one \ supplies a large class of
general relativistic cosmological solutions with Universal rotation.

Ni (2008;2009), has reported observations on a possible rotation of the
polarization of the cosmic background radiation, around 0.1 radians.As such
radiation was originated at the inception of the Universe, we tried to
estimate a possible angular speed or vorticity, by dividing 0.1 radians by the
age of the Universe , obtaining about 10$^{-19}$rad.s$^{-1}$.

.The numerical result is very close to the theoretical estimate, by Berman (2007),

\bigskip

$\omega\approx c/R=3.10^{-18}$rad.s$^{-1}.$

\bigskip

where $c$,$R$ represent the speed of light in vacuum, and the radius of the
causally related Universe.

\bigskip

\bigskip We must remember, as Berman and Gomide (2011) have pointed, that
their calculation deals with material particles, or, in the language of
General Relativity, non-null geodesics.The fact that the Universe may exhibit
a rotating state, can be understood by a simple fine-tuning argument---it
would be highly unprobable that the Universe could keep since birth a state of
no angular momentum at all.

\bigskip The value of Berman%
\'{}%
s rotation, fits with the Pioneers anomaly, which consists on decelerations
sufferred by Nasa space probes in non- closed curves, extending to outer
space.Thermal emission was cited as resolving the Pioneers anomaly, but it
does not explain the fly-bys, like Berman and Gomide (2011) did through the
present rotational theory.Worse, thermal emission is unable to explain why
elliptical orbiters do not decelerate accordingly.

\bigskip About this same numerical value of the angular speed is predicted
also in Godel%
\'{}%
s rotational model, but it is not \ an expanding one(see Adler, Bazin and
Schiffer,1975). In the next few years, the observational evidence may confirm
or not such rotation .

\bigskip

\bigskip Rotating metrics in General Relativity were first studied by Islam
(1985), but Cosmology was not touched upon. However, it would be necessary an
extreme perfect fine-tuning, in order to create the Universe without any
angular-momentum. The primordial Quantum Universe, is characterized by
dimensional combinations of the fundamental constants \ \textquotedblleft%
$c$\textquotedblright\ \ , \textquotedblleft$h$\textquotedblright\ \ \ and
\ \textquotedblleft$G$\textquotedblright\ \ \ respectively the speed of light
in vacuo, Planck's and Newton's gravitational constants. The natural angular
momentum of Planck's Universe, as it is called, is, then,
\ \ \textquotedblleft$h$\textquotedblright\ \ . It will be shown that the
angular momentum grows with the expanding Universe, but the corresponding
angular speed decreases with the scale-factor (or radius) of the Universe,
such being the reason for the difficulty in detection of this speed with
present technology. Notwithstanding, the so-called Pioneers' anomaly
(Anderson,2002), which is a deceleration verified in the Pioneers space-probes
launched by NASA more than thirty years ago, was attributed by Berman, to a
\textquotedblleft Machian\textquotedblright\ ubiquitous field of centripetal
accelerations, due to the rotation of the Universe. Berman's calculation
rested on the assumption that the zero-total energy of the Universe was a
valid result for the rotating case, but the proof was not supplied in that
paper (Berman, 2007b). By "proof", one thinks on the pseudotensor energy
calculations of General Relativity --- the best gravitational theory ever published.

\bigskip In his three best-sellers (Hawking, 1996; 2001; 2003), Hawking
describes inflation (Guth, 1981; 1998), as an accelerated expansion of the
Universe, immediately after the creation instant,while the Universe, as it
expands,borrows energy from the \ gravitational field to create more matter.
According to his \ description, the positive matter energy is exactly balanced
by the negative gravitational energy, so that the total energy is zero,and
that when \ the size of the Universe doubles, both the matter and
gravitational energies also double, keeping the total energy zero (twice
zero).Moreover, in the recent, next best-seller,Hawking and Mlodinow(2010)
comment that if it were not for the gravity interaction, one could not
\ validate a zero-energy Universe, and then, creation out of nothing would not
have happened.

\bigskip There are four methods,in GRT, to create \ rotations.Non-diagonal
metrics, like Kerr%
\'{}%
s, is one.The adoption of an imperfect fluid model, with vorticities, as in
Raychaudhuri%
\'{}%
s equation, is second.Third, you may follow the Godlowski et al (2004)
idea,and add to the scale-factor
\'{}%
s squared time derivative,$\dot{R}^{2}$a rotational term $(\omega R)^{2}.$On
the other hand, Berman (2008a; b) has shown that Robertson-Walker's metric, is
a particular, non-rotating case, of a general relativistic expanding and
rotating metric first developed by Gomide and Uehara (1981). The peculiarity
of the general metric is that instead of working with proper-time \ $\tau
$\ \ , one writes the field equations of General Relativity with a cosmic time
\ \ $t$\ \ \ related by:

\bigskip

$d\tau=(g_{00})^{1/2}dt$ \ \ \ \ \ \ \ \ \ \ \ \ \ \ \ \ \ \ \ \ \ \ \ \ , \ \ \ \ \ \ \ \ \ \ \ \ \ \ \ \ \ \ \ \ \ \ \ \ \ \ \ \ \ \ \ \ \ \ \ \ \ \ \ \ \ \ \ \ \ \ \ \ \ \ \ \ \ \ \ \ \ \ \ \ \ \ \ \ \ (1)

\bigskip

where,

\bigskip

$g_{00}=g_{00}(r,\theta,\phi,t)$ \ \ \ \ \ \ \ \ \ \ \ \ \ \ \ \ \ \ . \ \ \ \ \ \ \ \ \ \ \ \ \ \ \ \ \ \ \ \ \ \ \ \ \ \ \ \ \ \ \ \ \ \ \ \ \ \ \ \ \ \ \ \ \ \ \ \ \ \ \ \ \ \ \ \ \ \ \ \ \ \ \ \ \ \ (2)

\bigskip

It was seen that when one introduces a metric temporal coefficient \ $g_{00}%
$\ \ which is not constant, the new metric includes rotational effects. In
fact, we have a generalized Gaussian metric, because besides the fact that the
tri-space is orthogonal to the time-axis, the spatial part of the metric,
rotates as a whole, relative to this time axis. This is a new concept being
introduced in the theory.

\bigskip The present paper follows the steps of \ the semi-relativistic
treatment by Berman (2007), but this time, it is General relativistic, and we
shall find a Machian kind of solution.The general solution is to be found in
Berman and Gomide (2011a).

In a previous paper Berman (2009c) has calculated the energy of the
Friedman-Robertson-Walker's Universe, by means of pseudo-tensors, and found a
zero-total energy. \textbf{Our main task will be to show why the Universe is a
zero-total-energy entity, by means of pseudo-tensors, even when one chooses a
variable \ \ }$g_{00}$\textbf{\ \ \ such that the Universe also rotates, and
then, to show how General Relativity predicts a universal angular speed, and a
universal centripetal deceleration, numerically coincident with the observed
deceleration of the Pioneers space-probes. }The first calculation of this
kind, with the Gomide -Uehara generalization of RW%
\'{}%
s metric, was undertaken by Berman (1981), in his M.Sc. thesis, advised by the
present second author, but where the rotation of the Universe was not the
scope of the thesis.

\bigskip

\bigskip\bigskip The pioneer works of Nathan Rosen (Rosen, 1994), Cooperstock
and Israelit, (1995) , showing that the energy of the Universe is zero, by
means of calculations involving pseudotensors, and Killing vectors,
respectively, are here given a more simple approach. The energy of the
(non-rotating) Robertson-Walker's Universe is zero, (Berman, 2007;2009c).
Berman (1981) was the first author to work, in pseudotensor calculations for
the energy of Robertson-Walker's Universe. He made the calculations on which
the present paper rest, and, explicitly obtained the zero-total energy for a
closed Universe, by means of LL-pseudotensor, when Robertson-Walker's metric
was generalised by the introduction of a temporal-time-varying metric
coefficient. However, the present authors, were unaware, in the year 1981, of
the exact significance of their findings.

\bigskip

The zero-total-energy of the Roberston-Walker's Universe, and of any Machian
ones, have been shown by many authors (Berman 2006; 2006a; 2007; 2007a;
2007b). \bigskip It may be that the Universe might have originated from a
vacuum quantum fluctuation. In support of this view, we shall show that the
pseudotensor theory (Adler et al, 1975) points out to a null-energy for a
rotating Robertson-Walker's Universe. Some prior work is mentioned,(Berman
2006; 2006a; 2007; 2007a; 2007b; Rosen, 1995; York Jr, 1980; Cooperstock,
1994; Cooperstock and Israelit, 1995; Garecki,1995; Johri et al.,1995; Feng
and Duan,1996; Banerjee and Sen,1997; Radinschi,1999; Cooperstock and
Faraoni,2003). See also Katz (2006, 1985); Katz and Ori (1990); and Katz et al
(1997). Recent developments include torsion models (So and Vargas, 2006), and,
a paper by Xulu(2000).

The reason for the failure of non-Cartesian curvilinear coordinate energy
calculations through pseudotensors, resides in that curvilinear coordinates
carry non-null Christoffel symbols, even in Minkowski spacetime, thus
introducing inertial or fictitious fields that are interpreted falsely as
gravitational energy-carrying (false) fields.

Carmeli et al.(1990) listed four arguments against the use of Einstein%
\'{}%
s pseudotensor: \textbf{1.}the energy integral defines only an affine vector;
\textbf{2.}no angular-momentum is available; \textbf{3.} as it depends only on
the metric tensor and its first derivatives, it vanishes locally in a geodesic
system; \textbf{4.} due to the existence of a superpotential, which is related
to the total conserved pseudo-quadrimomentum, by means of a divergence, then
the values of \ the metric tensor, and its first derivatives, only matter, on
a surface around the volume of the mass-system.

\bigskip We shall argue below that, for the Universe, local and global Physics
blend together. The pseudo-momentum, is to be taken like the linear momentum
vector of Special Relativity, i.e., as an affine vector. In a previous paper
(Berman, 2009c), we stated that \ "if the Universe \ has some kind of
rotation, the energy-momentum calculation refers to a co-rotating observer".
Such being the case, we now go ahead for the actual calculations, involving
rotation. Birch (1982; 1983) cited inconclusive experimental data on a
possible rotation of the Universe, which was followed by a paper written by
Gomide, Berman and Garcia (1986).

\bigskip\bigskip

\bigskip

2.FIELD EQUATIONS FOR THE ROTATING AND EXPANDING METRIC

\bigskip

Consider first a temporal metric coefficient which depends only on \ $t$\ \ .
The line element becomes:\ 

\bigskip

$ds^{2}=-\frac{R^{2}(t)}{\left(  1+kr^{2}/4\right)  ^{2}}\left[  d\sigma
^{2}\right]  +g_{00}\left(  t\right)  $ $dt^{2}$
\ \ \ \ \ \ \ \ \ \ \ \ \ \ \ . \ \ \ \ \ \ \ \ \ \ \ \ \ \ \ \ \ \ \ \ \ \ \ \ \ \ \ \ \ \ \ \ \ \ \ \ \ \ \ \ \ \ \ \ \ \ (3)

\bigskip

The field equations, in \ General Relativity Theory (GRT) become:

\bigskip

$3\dot{R}^{2}=\kappa(\rho+\frac{\Lambda}{\kappa})g_{00}R^{2}-3kg_{00}$
\ \ \ \ \ \ \ \ \ \ \ \ \ \ \ \ \ \ \ \ \ , \ \ \ \ \ \ \ \ \ \ \ \ \ \ \ \ \ \ \ \ \ \ \ \ \ \ \ \ \ \ \ \ \ \ \ \ \ \ \ \ \ \ \ \ \ \ \ (4)

\bigskip

and,

\bigskip

$6\ddot{R}=-g_{00}\kappa\left(  \rho+3p-2\frac{\Lambda}{\kappa}\right)
R-3g_{00}\dot{R}$ $\dot{g}^{00}$ \ \ \ \ \ \ \ \ \ \ \ . \ \ \ \ \ \ \ \ \ \ \ \ \ \ \ \ \ \ \ \ \ \ \ \ \ \ \ \ \ \ \ \ \ \ \ \ \ \ \ \ \ \ (5)

\bigskip

Local inertial processes are observed through proper time, so that the
four-force is given by:

\bigskip

$F^{\alpha}=\frac{d}{d\tau}\left(  mu^{\alpha}\right)  =mg^{00}$ $\ddot
{x}^{\alpha}-\frac{1}{2}m$ $\dot{x}^{\alpha}\left[  \frac{\dot{g}_{00}}%
{g_{00}^{2}}\right]  $ \ \ \ \ \ \ \ \ \ \ \ \ \ \ . \ \ \ \ \ \ \ \ \ \ \ \ \ \ \ \ \ \ \ \ \ \ \ \ \ \ \ \ \ \ \ \ \ \ \ \ \ \ (6)

\bigskip

Of course, when \ $g_{00}=1$\ \ , the above equations reproduce conventional
Robertson-Walker's field equations.

\bigskip

We must mention that the idea behind Robertson-Walker's metric is the Gaussian
coordinate system. Though the condition \ \ $g_{00}=1$\ \ is usually adopted,
we must remember that, the resulting time-coordinate is meant as representing
proper time. If we want to use another coordinate time, we still keep the
Gaussian coordinate properties.

\bigskip

From the energy-momentum\ \ conservation equation, in the case of a uniform
Universe, \ we must have,

\bigskip

$\frac{\partial}{\partial x^{i}}\left(  \rho\right)  =\frac{\partial}{\partial
x^{i}}\left(  p\right)  =\frac{\partial}{\partial x^{i}}\left(  g_{00}\right)
=0$ \ \ \ \ \ \ \ \ \ \ \ \ \ ( \ $i=1,2,3$\ \ ) \ \ \ \ . \ \ \ \ \ \ \ \ \ \ \ \ \ \ \ \ \ \ \ \ \ \ \ \ \ \ (7)

\bigskip

The above is necessary in the determination of cosmic time, for a commoving
observer. We can see that the hypothesis \ (2) -- that \ $g_{00}$\ \ is only
time-varying -- is now validated.

\bigskip

In order to understand equation (6)\ , it is convenient to relate the
rest-mass $m$\ ,to \ an inertial mass \ $M_{i}$\ , with:

\bigskip

$M_{i}=\frac{m}{g_{00}}$\ \ \ \ \ \ \ \ \ \ \ \ \ \ .\ \ \ \ \ \ \ \ \ \ \ \ \ \ \ \ \ \ \ \ \ \ \ \ \ \ \ \ \ \ \ \ \ \ \ \ \ \ \ \ \ \ \ \ \ \ \ \ \ \ \ \ \ \ \ \ \ \ \ \ \ \ \ \ \ \ \ \ \ \ \ \ \ \ \ \ \ \ \ \ \ \ \ \ \ \ \ (8)

\bigskip

It can be seen that \ $M_{i}$\ \ represents the inertia of a particle, when
observed along cosmic time, i.e., coordinate time. In this case, we observe
that we have two acceleration terms, which we call,

\bigskip

\bigskip\ $a_{1}^{\alpha}=\ddot{x}^{\alpha}$\ \ \ \ \ \ \ \ \ \ \ \ \ \ \ , \ \ \ \ \ \ \ \ \ \ \ \ \ \ \ \ \ \ \ \ \ \ \ \ \ \ \ \ \ \ \ \ \ \ \ \ \ \ \ \ \ \ \ \ \ \ \ \ \ \ \ \ \ \ \ \ \ \ \ \ \ \ \ \ \ \ \ \ \ \ \ \ \ \ \ \ \ \ \ \ \ \ \ \ \ \ \ \ (9)

\bigskip

and,

\bigskip

\bigskip\ $a_{2}^{\alpha}=-\frac{1}{2g_{00}}\left(  \dot{x}^{\alpha}\dot
{g}_{00}\right)  $%
\ \ \ \ \ \ \ \ \ \ \ \ \ \ \ .\ \ \ \ \ \ \ \ \ \ \ \ \ \ \ \ \ \ \ \ \ \ \ \ \ \ \ \ \ \ \ \ \ \ \ \ \ \ \ \ \ \ \ \ \ \ \ \ \ \ \ \ \ \ \ \ \ \ \ \ \ \ \ \ \ \ \ \ \ \ \ \ (10)\bigskip

\bigskip

The first acceleration is linear; the second, resembles rotational motion, and
depends on \ $g_{00}$\ \ and its time-derivative.\ \ 

\bigskip

If we consider \ \ $a_{2}^{\alpha}$\ \ a centripetal acceleration, we conclude
that the angular speed \ \ $\omega$\ \ \ is given by,

\bigskip

$\omega=\frac{1}{2}\left(  \frac{\dot{g}_{00}}{g_{00}}\right)  $%
\ \ \ \ \ \ \ \ \ \ \ \ \ \ \ \ \ . \ \ \ \ \ \ \ \ \ \ \ \ \ \ \ \ \ \ \ \ \ \ \ \ \ \ \ \ \ \ \ \ \ \ \ \ \ \ \ \ \ \ \ \ \ \ \ \ \ \ \ \ \ \ \ \ \ \ \ \ \ \ \ \ \ \ \ \ \ \ \ \ \ \ \ \ (11)

\bigskip

By comparison between the usual \ $\tau$ -- metric, and the field equations in
the \ $t$\ -- metric, we are led to conclude that the conventional energy
density \ $\rho$\ \ \ and cosmic pressure \ $p$\ \ \ are transformed into
\ \ $\bar{\rho}$\ \ \ and \ \ $\bar{p}$\ \ , where:

\bigskip

$\bar{\rho}=g_{00}\left(  \rho+\frac{\bar{\Lambda}}{\kappa}\right)
$\ \ \ \ \ \ \ \ \ \ \ \ \ \ \ \ , \ \ \ \ \ \ \ \ \ \ \ \ \ \ \ \ \ \ \ \ \ \ \ \ \ \ \ \ \ \ \ \ \ \ \ \ \ \ \ \ \ \ \ \ \ \ \ \ \ \ \ \ \ \ \ \ \ \ \ \ \ \ \ \ \ \ \ \ \ \ \ \ (12)

\bigskip

and,

\bigskip

$\bar{p}=g_{00}\left(  p-\frac{\bar{\Lambda}}{\kappa}\right)  $%
\ \ \ \ \ \ \ \ \ \ \ \ \ \ \ \ . \ \ \ \ \ \ \ \ \ \ \ \ \ \ \ \ \ \ \ \ \ \ \ \ \ \ \ \ \ \ \ \ \ \ \ \ \ \ \ \ \ \ \ \ \ \ \ \ \ \ \ \ \ \ \ \ \ \ \ \ \ \ \ \ \ \ \ \ \ \ \ \ (13)

\bigskip

We plug back into the field equations, and find,

\bigskip

$\bar{\Lambda}=\Lambda-\frac{3}{2\kappa}\left(  \frac{\dot{R}}{R}\right)
\dot{g}^{00}$ \ \ \ \ \ \ \ \ \ \ \ \ \ . \ \ \ \ \ \ \ \ \ \ \ \ \ \ \ \ \ \ \ \ \ \ \ \ \ \ \ \ \ \ \ \ \ \ \ \ \ \ \ \ \ \ \ \ \ \ \ \ \ \ \ \ \ \ \ \ \ \ \ \ \ \ \ \ \ \ \ \ \ (14)

\bigskip

For a time-varying angular speed, considering an arc $\phi$\ , so that,

\bigskip

$\omega(t)=\frac{d\phi}{dt}=\dot{\phi}$\ \ \ \ \ \ \ \ \ \ \ \ \ \ \ \ \ , \ \ \ \ \ \ \ \ \ \ \ \ \ \ \ \ \ \ \ \ \ \ \ \ \ \ \ \ \ \ \ \ \ \ \ \ \ \ \ \ \ \ \ \ \ \ \ \ \ \ \ \ \ \ \ \ \ \ \ \ \ \ \ \ \ \ \ \ \ \ \ \ \ \ \ (15)

\bigskip

we find, from (11),

\bigskip

$g_{00}=Ce^{2\phi(t)}$ \ \ \ \ \ \ \ . \ \ \ ( \ $C$\ \ = constant )\ \ \ \ \ \ \ \ \ \ \ \ \ \ \ \ \ \ \ \ \ \ \ \ \ \ \ \ \ \ \ \ \ \ \ \ \ \ \ \ \ \ \ \ \ \ \ \ \ \ \ \ \ \ \ \ \ \ (16)

\bigskip

Returning to (14), we find,

\bigskip

$\bar{\Lambda}=\Lambda+\frac{3}{\kappa C}\left(  \frac{\dot{R}}{R}\right)
\omega e^{-2\phi(t)}$ \ \ \ \ \ \ \ \ \ \ \ \ \ . \ \ \ \ \ \ \ \ \ \ \ \ \ \ \ \ \ \ \ \ \ \ \ \ \ \ \ \ \ \ \ \ \ \ \ \ \ \ \ \ \ \ \ \ \ \ \ \ \ \ \ \ \ \ \ \ \ \ \ \ \ \ \ (17)

\bigskip

\bigskip This completes our solution.

\bigskip The case where \ $g_{00}$\ depends also\ on \ $r,\theta$\ \ and
\ $\phi$\ \ \ was considered also by Berman (2008b)\ and does not differ
qualitatively from the present analysis, so that, we refer the reader to that paper.

\bigskip

\bigskip

\bigskip3.ENERGY OF THE ROTATING EVOLUTIONARY UNIVERSE

\bigskip

Even in popular Science accounts(Hawking,1996;2001;2003\bigskip;--- and
Moldinow,2010; Guth,1998), it has been generally accepted that the Universe
has zero-total energy. The first such claim, seems to be due to
Feynman(1962-3). Lately, Berman(2006, 2006 a) has proved this result by means
of simple arguments involving Robertson-Walker's metric for any value of the
tri-curvature ( $0,-1,1$ ).

The pseudotensor \ $t_{\nu}^{\mu}$\ , also called Einstein's pseudotensor, is
such that, when summed with the energy-tensor of matter \ $T_{\nu}^{\mu}$\ \ ,
gives the following conservation law:\ \ \ 

\bigskip

$\left[  \sqrt{-g}\left(  T_{\nu}^{\mu}+t_{\nu}^{\mu}\right)  \right]  ,_{\mu
}=0$ \ \ \ \ \ \ \ \ \ \ \ \ \ \ \ \ \ \ \ \ .\ \ \ \ \ \ \ \ \ \ \ \ \ \ \ \ \ \ \ \ \ \ \ \ \ \ \ \ \ \ \ \ \ \ \ \ \ \ \ \ \ \ \ \ \ \ \ \ \ (18)

\bigskip

In such case, the quantity

\bigskip

$P_{\mu}=\int\left\{  \sqrt{-g}\left[  T_{\mu}^{0}+t_{\mu}^{0}\right]
\right\}  $ $d^{3}x$ \ \ \ \ \ \ \ \ \ \ \ \ \ \ \ \ \ \ \ \ ,\ \ \ \ \ \ \ \ \ \ \ \ \ \ \ \ \ \ \ \ \ \ \ \ \ \ \ \ \ \ \ \ \ \ \ \ \ \ \ \ \ (19)

\bigskip

is called the general-relativistic generalization of the energy-momentum
four-vector of special relativity (Adler et al, 1975).

\bigskip

It can be proved that \ $P_{\mu}$\ \ \ is conserved when:

\bigskip

a) \ $T_{\nu}^{\mu}\neq0$\ \ only in a finite part of space; \ \ \ \ \ \ and,\ \ 

b) \ $g_{\mu\nu}\rightarrow\eta_{\mu\nu}$ \ when we approach infinity, where
\ $\eta_{\mu\nu}$\ \ is the Minkowski metric tensor.

\bigskip

However, there is no reason to doubt that, even if the above conditions were
not fulfilled, we might eventually get a constant \ $P_{\mu}$\ , because the
above conditions are sufficient, but not strictly necessary. We hint on the
plausibility of other conditions, instead of \ \ a) \ and \ \ b) \ \ above.

\bigskip

Such a case will occur, for instance, when we have the integral in (19)\ \ is
equal to zero.

\bigskip

For our generalised metric, we get exactly this result, because, from Freud's
(1939) formulae, there exists a super-potential, (Papapetrou, 1974):

\bigskip

$_{F}U_{\lambda}^{\mu\nu}=\frac{g_{\lambda\alpha}}{2\sqrt{-g}}(\bar{g}%
^{\mu\alpha}\bar{g}^{\nu\beta}-\bar{g}^{\nu\alpha}\bar{g}^{\mu\beta
}),_{_{\beta}}$ \ \ \ \ \ \ \ \ \ \ \ \ \ \ \ \ \ \ \ \ ,

\bigskip

where the bars over the metric coefficients imply that they are multiplied by
\ \ $\sqrt{-g}$\ \ , and such that,

\bigskip

$\kappa\sqrt{-g}(T_{\lambda}^{\rho}+t_{\lambda}^{\rho})=$\ $_{F}U_{\lambda
}^{\rho\sigma},_{\sigma}$ \ \ \ \ \ \ \ \ \ \ \ \ \ \ \ \ \ \ \ \ ,

\bigskip

thus finding, after a brief calculation, for the rotating Robertson-Walker's metric,

\bigskip

$P_{\lambda}=0$ \ \ \ \ \ \ \ \ \ \ \ \ \ \ \ \ \ \ \ \ \ \ \ \ \ \ \ \ \ \ \ \ \ .

The above result, with von Freud's superpotential, which yields Einstein's
pseudotensorial results, points to a zero-total energy Universe, even when the
metric is endowed with a varying metric temporal coefficient .

\bigskip

A similar result would be obtained from Landau-Lifshitz pseudotensor
(Papapetrou, 1974), where we have:

\bigskip

$P_{LL}^{\nu}=\int(-g)\left[  T^{\nu0}+t_{L}^{\nu0}\right]  $ $\ d^{3}x$
\ \ \ \ \ \ \ \ \ \ \ \ \ \ \ \ , \ \ \ \ \ \ \ \ \ \ \ \ \ \ \ \ \ \ \ \ \ \ \ \ \ \ \ \ \ \ \ \ \ \ \ \ \ \ \ \ \ \ \ \ \ (20)

\bigskip

where,

\bigskip

\bigskip$\kappa\sqrt{-g}(T^{\mu\rho}+\tilde{t}^{\mu\rho})=$\ $\tilde{U}%
^{\mu\rho\sigma},_{\sigma}$ \ \ \ \ \ \ \ \ \ \ \ \ \ \ \ \ \ \ \ \ ,

\bigskip

and, \ \ \ $\tilde{U}^{\mu\rho\sigma}=\bar{g}_{{}}^{\lambda\mu}$
$_{F}U_{\lambda}^{\rho\sigma}$ \ \ \ \ \ \ \ \ \ \ \ \ \ \ \ \ \ \ \ \ ,

\bigskip

\bigskip A short calculation shows that, for the rotating metric,too, we keep
valid the result,

\bigskip

$P_{LL}^{\nu}=0$ \ \ \ \ \ \ \ \ \ \ \ \ \ \ \ \ ( $\nu=0,1,2,3$%
\ \ )\ \ \ \ \ \ \ \ . \ \ \ \ \ \ \ \ \ \ \ \ \ \ \ \ \ \ \ \ \ \ \ \ \ \ \ \ \ \ \ \ \ \ \ \ \ \ (21)\ \ \ \ \ \ \ \ \ 

\bigskip

\bigskip Other superpotentials would also yield the same zero results. A
useful source for the main superpotentials in the market, is the paper by
Aguirregabiria et al. (1996).

\bigskip

\bigskip The equivalence principle, says that at any location, spacetime is
\ (locally) flat, and a geodesic coordinate system may be constructed,
\textit{where the Christoffel symbols are null. The pseudotensors are, then,
at each point, null. But now remember that our old Cosmology requires a
co-moving observer at each point}. \textit{It is this co-motion that is
associated with the geodesic system, and, as RW%
\'{}%
s metric is homogeneous and isotropic, for the co-moving observer, the
zero-total energy density result, is repeated from point to point, all over
spacetime. Cartesian coordinates are needed, too, because curvilinear
coordinates are associated with fictitious or inertial forces, which would
introduce inexistent accelerations that can be mistaken \ additional
gravitational fields (i.e.,that add to the real energy). Choosing Cartesian
coordinates is not analogous to the use of center of mass \ \ frame in
Newtonian theory, but the null results for the spatial components of the
pseudo-quadrimomentum show compatibility. }

\bigskip\bigskip

\bigskip4.AN ALTERNATIVE DERIVATION

Though so many researchers \ have dealt with the energy of the Universe, our
present original solution involves rotation. We may paraphrase a previous
calculation, provided that we work with proper time \ \ $\tau$\ \ \ \ instead
of coordinate time \ \ $t$\ \ \ \ (Berman, 2009c). Then, the rotation of the
Universe will be automatically included. We shall now consider, first, why the
Minkowski metric represents a null energy Universe . Of course, it is empty.
But, why it has zero-valued energy? We resort to the result of Schwarzschild%
\'{}%
s metric, (Adler et al., 1975), whose total energy is,

\bigskip

$E=Mc%
{{}^2}%
-\frac{GM%
{{}^2}%
}{2R}$ \ \ \ \ \ \ \ \ \ \ \ \ \ \ \ \ \ \ \ \ \ \ . \ \ \ \ \ \ \ \ \ \ \ \ \ \ \ \ \ \ \ \ \ \ \ \ \ \ \ \ \ \ \ \ \ \ \ \ \ \ \ \ \ \ \ \ \ \ \ \ \ \ \ \ \ \ \ 

\bigskip

If \ \ \ $M=0$ \ \ , \ the energy is zero,too. But when we write Schwarzschild%
\'{}%
s metric, and make \ \ the mass become zero, we obtain Minkowski metric, so
that we got the zero-energy result. Any flat RW%
\'{}%
s metric, can be reparametrized as Minkowski%
\'{}%
s; or,for closed and open Universes, a superposition of such cases
(Cooperstock and Faraoni,2003; Berman, 2006; 2006a).

\bigskip

Now, the energy of the Universe, can be calculated at constant time coordinate
\ \ $\tau$ \ . In particular, the result would be the same as when
\ \ $\tau\rightarrow\infty$ \ \ , or, even when \ \ $\tau\rightarrow0$ \ \ .
\bigskip Arguments for initial null energy come from Tryon(1973), and Albrow
(1973).More recently, we recall the quantum fluctuations of Alan Guth%
\'{}%
s inflationary scenario (Guth,1981;1998). Berman (see for instance,2008c),
gave the Machian picture of the Universe, as being that of a zero energy .
Sciama%
\'{}%
s inertia theory results also in a zero-total energy Universe (Sciama, 1953;
Berman, 2008d;2009e).

\bigskip

Consider the possible solution for the rotating case. We \ work with the
$\tau$-metric, so that we keep formally the RW%
\'{}%
s metric in an accelerating Universe. The scale-factor assumes a power-law ,
as in constant deceleration parameter models (Berman,1983;---and Gomide,1988),

\bigskip

$R=(mD\tau)^{1/m}$ \ \ \ \ \ \ \ \ \ \ \ \ \ \ \ \ \ \ \ \ \ \ \ \ \ \ \ \ \ \ \ \ \ \ \ \ \ \ \ \ \ \ ,\ \ \ \ \ \ \ \ \ \ \ \ \ \ \ \ \ \ \ \ \ \ \ \ \ \ \ \ \ \ \ \ \ \ \ \ \ (22)

\bigskip

where, $\ m$ $\ $, $\ \ D=$ \ constants, \ and,

\bigskip

$m=q+1>0$
\ \ \ \ \ \ \ \ \ \ \ \ \ \ \ \ \ \ \ \ \ \ \ \ \ \ \ \ \ \ \ \ \ \ \ \ \ \ \ \ \ \ \ \ \ ,
\ \ \ \ \ \ \ \ \ \ \ \ \ \ \ \ \ \ \ \ \ \ \ \ \ \ \ \ \ \ \ \ \ (23)

\bigskip where \ $q$ \ \ \ is the deceleration parameter.

For a perfect fluid energy tensor, and a perfect gas equation of state, cosmic
pressure and energy density obey the following energy-momentum conservation
law, \bigskip(Berman, 2007, 2007a),

\bigskip

$\dot{\rho}=-3H(\rho+p)$
\ \ \ \ \ \ \ \ \ \ \ \ \ \ \ \ \ \ \ \ \ \ \ \ \ \ \ \ \ \ \ \ , \ \ \ \ \ \ \ \ \ \ \ \ \ \ \ \ \ \ \ \ \ \ \ \ \ \ \ \ \ \ \ \ \ \ \ \ \ \ \ \ \ \ (24)

\bigskip

where, only in this Section, overdots stand for $\tau$-derivatives .Let us have,

\bigskip

$p=\alpha\rho$ \ \ \ \ \ \ \ \ \ \ ( $\ \alpha=$ \ constant larger than $-1$
\ \ ) \ \ \ .\ \ \ \ \ \ \ \ \ \ \ \ \ \ \ \ \ \ \ \ \ \ \ \ \ \ (25)

\bigskip

On solving the differential equation, we find, for any \ \ $k=0$ \ \ , \ \ $1$
\ \ \ , \ \ $-1$ \ \ ,that,

\bigskip

$\rho=\rho_{0}\tau^{-\frac{3(1+\alpha)}{m}}$
\ \ \ \ \ \ \ \ \ \ \ \ (\ \ $\rho_{0}=$ \ \ constant)\ \ \ \ \ \ \ \ . \ \ \ \ \ \ \ \ \ \ \ \ \ \ \ \ \ \ \ \ \ \ \ \ \ \ \ \ \ \ \ \ \ \ (26)\ 

\ \ \ \ \ \ 

When \ \ $\tau\rightarrow\infty$ \ \ , from (26) we see that the energy
density becomes zero, and we retrieve an "empty" Universe, or, say, again, the
energy is zero. However, this energy density is for the matter portion, but
nevertheless, as in this case, $\ \ R\rightarrow\infty$ \ \ , all masses are
infinitely far from each others, so that the gravitational inverse-square
interaction is also null. The total energy density is null, and, so, the total
energy. Notice that the energy-momentum conservation equation does not change
even if we add a cosmological constant density, because we may subtract an
equivalent amount in pressure, and \ equation (24) remains the same. The
constancy of the energy, leads us to consider the zero result at infinite
time, also valid at any other instant.

\bigskip

We refer to Berman (2006; 2006a) for another alternative proof of the
zero-energy Universe. If we took \ \ \ $\tau$\ \ \ \ \ instead of
\ \ \ \ \ $t$\ \ \ \ \ \ , these references would provide the zero result also
for the rotational case.

\bigskip

\bigskip

5.PIONEERS ANOMALY REVISITED

\bigskip

Einstein's field equations (4) and (5) above, can be obtained, when
\ $g_{00}=$\ \ constant, \ through the mere assumptions of conservation of
energy (equation 4)\ \ and thermodynamical balance of energy (equation 5), as
was pointed out by Barrow (1988). The latter is also to be regarded as a
definition of cosmic pressure, as the volume derivative of energy with
negative sign \ \ ( \ \ $p=-\frac{d(\rho V)}{dV}$\ \ \ \ ) .

\bigskip

Now, let us consider a time-varying $g_{00}$. We may write the energy (in
fact, the "energy-density")-- equation, as follows:

\bigskip

\bigskip$\frac{3\dot{R}^{2}}{g_{00}}-\kappa(\rho+\frac{\Lambda}{\kappa}%
)R^{2}=-3k=$ \ \ constant\ \ \ \ \ \ \ \ \ \ \ \ \ \ \ \ \ \ \ . \ \ \ \ \ \ \ \ \ \ \ \ \ \ \ \ \ \ \ \ \ \ \ \ \ \ \ \ \ \ \ \ \ (27)

\bigskip

The r.h.s. stands for a constant. We can regard the l.h.s. as the a sum of
constant terms, thus finding a possible solution of the field equations, such
that each term in the l.h.s. of (27) remains constant. For example, let us consider,

\bigskip

$\rho=\rho_{0}R^{-2}$
\ \ \ \ \ \ \ \ \ \ \ \ \ \ \ \ \ \ \ \ \ \ \ \ \ \ \ \ , \ \ \ \ \ \ \ \ \ \ \ \ \ \ \ \ \ \ \ \ \ \ \ \ \ \ \ \ \ \ \ \ \ \ \ \ \ \ \ \ \ \ \ \ \ \ \ \ \ \ \ \ \ \ \ \ \ \ \ \ \ \ \ \ \ \ \ \ (28)

\bigskip

$\Lambda=\Lambda_{0}R^{-2}$
\ \ \ \ \ \ \ \ \ \ \ \ \ \ \ \ \ \ \ \ \ \ \ \ \ \ \ , \ \ \ \ \ \ \ \ \ \ \ \ \ \ \ \ \ \ \ \ \ \ \ \ \ \ \ \ \ \ \ \ \ \ \ \ \ \ \ \ \ \ \ \ \ \ \ \ \ \ \ \ \ \ \ \ \ \ \ \ \ \ \ \ \ \ \ \ (29)

\bigskip

$g_{00}=3\gamma^{-1}\dot{R}^{2}$
\ \ \ \ \ \ \ \ \ \ \ \ \ \ \ \ \ \ \ \ \ \ \ ,
\ \ \ \ \ \ \ \ \ \ \ \ \ \ \ \ \ \ \ \ \ \ \ \ \ \ \ \ \ \ \ \ \ \ \ \ \ \ \ \ \ \ \ \ \ \ \ \ \ \ \ \ \ \ \ \ \ \ \ \ \ \ \ \ \ \ \ \ (30)\bigskip

\bigskip

where, \ \ \ \ $\rho_{0}$\ \ \ , \ $\Lambda_{0}$ \ \ and \ \ $\gamma
$\ \ \ \ are non-zero constants.

\bigskip

When we plug the above solution to the cosmic pressure equation (5), we find
that it is automatically satisfied provided that the following conditions hold,

\bigskip

$2\Lambda_{0}=\kappa\rho_{0}(1+3\alpha)$
\ \ \ \ \ \ \ \ \ \ \ \ \ \ \ \ \ \ \ \ \ \ , \ \ \ \ \ \ \ \ \ \ \ \ \ \ \ \ \ \ \ \ \ \ \ \ \ \ \ \ \ \ \ \ \ \ \ \ \ \ \ \ \ \ \ \ \ \ \ \ \ \ \ \ \ \ \ \ \ (31)

\bigskip

$p=\alpha\rho$ \ \ \ \ \ \ \ \ \ \ \ \ \ $(\alpha=$ constant$)$ \ \ \ \ \ ,\ \ \ \ \ \ \ \ \ \ \ \ \ \ \ \ \ \ \ \ \ \ \ \ \ \ \ \ \ \ \ \ \ \ \ \ \ \ \ \ \ \ \ \ \ \ \ \ \ \ \ \ \ \ \ \ \ (32)

\bigskip

and,

$\bigskip$

$\gamma=\kappa\rho_{0}+\Lambda_{0}-3k$ \ \ \ \ \ \ \ \ \ \ \ \ \ \ \ \ \ \ \ \ \ \ \ \ \ .\ \ \ \ \ \ \ \ \ \ \ \ \ \ \ \ \ \ \ \ \ \ \ \ \ \ \ \ \ \ \ \ \ \ \ \ \ \ \ \ \ \ \ \ \ \ \ \ \ \ \ \ \ \ \ \ \ \ (32a)

\bigskip

As we found a general-relativistic solution, so far, we are entitled to the
our previous general relativistic angular speed formula (11), to which we plug
\ our solution (30), to wit,

\bigskip

$\omega=\frac{\ddot{R}}{\dot{R}}=H+\frac{\dot{H}}{H}$ \ \ \ \ \ .

\bigskip For the power-law solution of the last Section,

\bigskip$H=\frac{1}{mt}$ \ \ \ \ \ \ ,

so that,

$\bigskip$

$\omega=-\frac{q}{mt}\approx t^{-1}$ \ \ \ \ \ \ \ ,

\bigskip where we roughly estimated the present deceleration paramenter as
$-1/2$, while, the centripetal acceleration,

\bigskip

$a=-\omega^{2}R\approx-t^{-2}R\simeq8.10^{-8}$ cm.s$^{-2}.$

\bigskip

Notice that the same result would follow from a scale-factor varying linearly
with time. This is the sort of scale-factor associated with the Machian
Universe. In fact,the field equations that we had (equations (4) and (5)),
were not enough in order to determine the exact form of the scale-factor,
because we had an extra-unknown term, the temporal metric coefficient. When we
advance a given equation of state, the original RW%
\'{}%
s field equations, with constant $g_{00}$,may determine the scale-factor%
\'{}%
s formula. Just to remember, our solution is a particular one.

\bigskip This is a general relativistic result. It matches Pioneers anomalous deceleration.

In an Appendix to this Section, we go ahead with the alternative calculation
with a simple naive Special Relativistic - Machian analysis, as had been made
in Berman(2007b).

\bigskip\ 

\bigskip APPENDIX TO THIS SECTION

\bigskip

As we now have the pseudo-tensorial zero-total energy result, for rotation
plus expansion, we might write in terms of \ elementary Physics, a possible
energy of the Universe equation, composed of the inertial term of Special
Relativity,\ \ $Mc^{2}$\ \ \ , the potential self-energy \ \ $-\frac{GM^{2}%
}{2R}$\ \ \ , and the cosmological "constant"\ energy, \ \ \ $\frac{\Lambda
}{\kappa}(\frac{4}{3}\pi R^{3})$\ \ \ \ , and not forgetting rotational
energy, \ \ $\frac{1}{2}I\omega^{2}$\ \ \ , where \ \ \ $I$\ \ \ stands for
the moment of inertia of a "sphere"\ \ of radius \ \ \ \ $R$\ \ \ \ \ and mass
\ \ \ $M$\ \ \ . The energy equation is equated to zero, i.e.,

\bigskip

$0=Mc^{2}-\frac{GM^{2}}{2R}+\frac{\Lambda}{\kappa}(\frac{4}{3}\pi R^{3}%
)+\frac{1}{2}I\omega^{2}$ \ \ \ \ \ \ \ \ \ \ \ \ \ \ . \ \ \ \ \ \ \ \ \ \ \ \ \ \ \ \ \ \ \ \ \ \ \ \ \ \ \ \ \ \ \ \ \ \ \ \ \ (33)

\bigskip

It must be remembered that \ \ \ \ $R$\ \ \ \ \ is a time-increasing function,
while the total-zero energy result must be time-invariant, so that the
principle of energy conservation be valid. A close analysis shows that the
above conditions can be met by solutions (28) and (29), which were derived or
induced from the general relativistic equations. When we plug the inertia moment,

\bigskip

$I=\frac{2}{5}MR^{2}$ \ \ \ \ \ \ \ \ \ \ \ \ \ \ \ \ \ \ , \ \ \ \ \ \ \ \ \ \ \ \ \ \ \ \ \ \ \ \ \ \ \ \ \ \ \ \ \ \ \ \ \ \ \ \ \ \ \ \ \ \ \ \ \ \ \ \ \ \ \ \ \ \ \ \ \ \ \ \ \ \ \ \ \ \ \ \ \ \ \ \ \ \ \ \ (34)

\bigskip

we need also to consider the following Brans-Dicke generalised relations,

\bigskip

$\frac{GM}{c^{2}R}=\Gamma=$ \ constant \ \ \ \ \ \ \ \ \ \ \ \ \ \ \ , \ \ \ \ \ \ \ \ \ \ \ \ \ \ \ \ \ \ \ \ \ \ \ \ \ \ \ \ \ \ \ \ \ \ \ \ \ \ \ \ \ \ \ \ \ \ \ \ \ \ \ \ \ \ \ \ \ \ \ \ \ \ \ (35)

\bigskip

and,

\bigskip

$\omega=\frac{c}{R}$ \ \ \ \ \ \ \ \ \ \ \ \ \ \ \ \ \ \ \ \ \ \ . \ \ \ \ \ \ \ \ \ \ \ \ \ \ \ \ \ \ \ \ \ \ \ \ \ \ \ \ \ \ \ \ \ \ \ \ \ \ \ \ \ \ \ \ \ \ \ \ \ \ \ \ \ \ \ \ \ \ \ \ \ \ \ \ \ \ \ \ \ \ \ \ \ \ \ \ \ \ (36)

\bigskip

If we calculate the centripetal acceleration corresponding to the above
angular speed, we find, for the present Universe, with $R\approx10^{28}$cm and
$\ c\simeq3.10^{10}$cm.s$^{-2}$ \ \ ,

\bigskip

$a_{cp}=-\omega^{2}R\cong-8.10^{-8}cm/s^{2}$
\ \ \ \ \ \ \ \ \ \ \ \ \ \ \ \ \ \ \ . \ \ \ \ \ \ \ \ \ \ \ \ \ \ \ \ \ \ \ \ \ \ \ \ \ \ \ \ \ \ \ \ \ \ \ \ \ \ \ \ \ \ \ \ \ \ \ \ (37)

\bigskip

This value matches the observed experimentally deceleration of the NASA
Pioneers' space-probes.

\bigskip

We observe that the Machian picture above is understood to be valid for any
observer in the Universe, i.e., the center of the "ball" coincides with any
observer; the "Machian" centripetal acceleration should be felt by any
observed point in the Universe subject to observation from any other location.

\bigskip We solve also other mistery concerning Pioneers anomaly. It has been
verified experimentally, that those space-probes in closed (elliptical) orbits
do not decelerate anomalously, but only those in hyperbolic flight. The
solution of this other enigma is easy, according to our view. The elliptical
orbiting trajectories are restricted to our local neighborhood, and do not
acquire cosmological features, which are necessary to qualify for our Machian
analysis, which centers on cosmological ground. But hyperbolic motion is not
bound \ by the Solar system, and in fact those orbits extend to infinity, thus
qualifying themselves to suffer the cosmological Machian deceleration.Thermal
emission may solve the first Pioneer anomaly, but it does not solve the
spin-down, nor the fly-bys in gravity assists.It is not clear why, thermal
emission did not cause decelerations in elliptical orbiters.Rotation of the
Universe solves all the three (Berman and Gomide,2011a).

\bigskip

\bigskip6.FINAL COMMENTS AND DISCUSSION

\bigskip

Someone has made very important criticisms on our work.First, he says why do
not the planets in the solar system show the calculated deceleration on the
Pioneers?The reason is that elliptical orbits are closed, and localized.You do
not feel the expansion of the universe in the sizes of the orbits either.In
General Relativity books, authors make this explicit.You do not include Hubble%
\'{}%
s expansion in Schwarzschild%
\'{}%
s metric.But, those space probes that undergo hyperbolic motion, which orbits
extend towards infinity, they acquire cosmological characteristics, like, the
given P.A. deceleration.Second objection, there are important papers which
resolve the P.A. with non-gravitational Physics.The answer,--- that is OK, we
have now alternative explanations.This does not preclude ours.Third,
cosmological reasons were discarded, including rotation of the Universe.The
problem is that those discarded cosmologies, did not employ the correct
metric.For instance, they discarded rotation by examining Godel model, which
is non expanding, and with a strange metric. The kind of metric we employ now,
or the one that we employed in the rotational case, were not discarded or
discussed by the authors cited by this objecter. Then, the final question, is
how come that a well respected author dismissed planetary Coriolis forces
induced by rotation of distant masses, by means of the constraints in the
solar system.Our answer is that, beside what we answered above, he needs to
consider Mach%
\'{}%
s Principle on one side, and the theoretical meaning of vorticities, because
one is not speaking in a center or an axis of rotation or so.When we say, in
Cosmology, that the Universe rotates, we mean that there is a field of
vorticities,just that.The whole idea is that Cosmology does not enter the
Solar System except for non-closed orbits that extend to outer space.We ask
the reader \ to check Mach%
\'{}%
s Principle, because in some formulations of this principle, rotation is in
fact a \textit{forbidden affaire}.

\bigskip Another one pointed out a different "problem". He objects, that the
angular speed \ formula of ours, is coordinate dependent.Now, when you choose
a specific metric, you do it thinking about the kind of problem you have to
tackle.After you choose the convenient metric, you forget tensor calculus, and
you work with coordinate-dependent relations.They work only for the given
metric, of course.

\bigskip

\bigskip We have obtained a zero-total energy proof for a rotating expanding
Universe. The zero result for the spatial components of the
energy-momentum-pseudotensor calculation, are equivalent to the choice of a
center of Mass reference system in Newtonian theory, likewise the use of
comoving observers in Cosmology. It is with this idea in mind, that we are led
to the energy calculation, yielding zero total energy, for the Universe, as an
acceptable result: we are assured that we chose the correct reference system;
this is a response to the criticism made by some scientists which argue that
pseudotensor calculations depend on the reference system, and thus, those
calculations are devoid of physical meaning.

\bigskip

Related conclusions by Berman should be consulted (see all Berman's references
at the end of this article). As a bonus, we can assure that there was not an
initial infinite energy density singularity, because attached to the
zero-total energy conjecture, there is a zero-total energy-density result, as
was pointed by Berman elsewhere (Berman, 2008).The so-called total energy
density of the Universe, which appears in some textbooks, corresponds only to
the non-gravitational portion, and the zero-total energy density results when
we subtract from the former, the opposite potential energy density.

\bigskip

As Berman(2009d; f) shows, we may say that the Universe is \emph{singularity
-free}, and was created \emph{ab-nihilo}, nor there is zero-time infinite
energy-density singularity.

\bigskip Paraphrasing Dicke (1964; 1964a), it has been shown the many faces of
Dirac's LNH, as many as there are about Mach's Principle. In face of \ modern
Cosmology, the naif theory of Dirac is a foil for theoretical discussion on
the foundations of this branch of Physical theory. The angular speed found by
us,(Berman,2010;2009a), matches results by G\"{o}del (see Adler et al., 1975),
Sabbata and Gasperini (1979), and Berman (2007b, 2008b, c).

\bigskip

\bigskip Rotation of the Universe and zero-total energy were verified for
Sciama's linear theory, which has been expanded, through the analysis of
radiating processes, by one of the present authors (Berman,
2008d;2009e).There,we found Larmor's power formula, in the gravitational
version, leads to the correct constant power relation for the Machian
Universe. However, we must remember that in local Physics, General Relativity
deals with quadrupole radiation, while Larmor is a dipole formula; for the
Machian Universe the resultant constant power is basically the same, either
for our Machian analysis or for the Larmor and general relativistic formulae.

\bigskip

Referring to rotation, it could be argued that cosmic microwave background
radiation \ deals with null geodesics, while Pioneers' anomaly, for instance,
deals with time-like geodesics. In favor of evidence on rotation, we remark
neutrinos' spin, parity violations, the asymmetry between matter and
anti-matter, left-handed DNA-helices, the fact that humans and animals alike
have not symmetric bodies, the same happening to molluscs.And, of course, the
results of the rotation of the polarization of CMBR.

\bigskip

We predict that chaotic phenomena and fractals, rotations in galaxies and
clusters, may provide clues on possible left handed preference through the Universe.

\bigskip

Berman and Trevisan (2010) have remarked that creation out-of-nothing seems to
be supported by the zero-total energy calculations. Rotation was now included
in the derivation of the zero result. We could think that the Universes are
created in pairs, the first one (ours), has negative spin and positive matter;
the second member of the pair, would have negative matter and positive spin:
for the ensemble of the two Universes, the total mass would always be zero;
the total spin, too. The total energy (twice zeros) is also zero.Our
framework, is the only one to solve the fly-by anomaly altogether, and
explains why elliptical orbiters do not decelerate.

{\LARGE \bigskip}

\bigskip

ACKNOWLEDGEMENTS

\bigskip

The authors thank Marcelo Fermann Guimar\~{a}es, Nelson Suga, Mauro Tonasse,
Antonio F. da F. Teixeira, and for the encouragement by Albert, Paula, and
Luiza Mitiko Gomide.

\bigskip

\bigskip

\bigskip

\bigskip

REFERENCES AND RELATED BIBLIOGRAPHY

\bigskip

Adler, R.J.; Bazin, M.; Schiffer, M. (1975) - \ \textit{Introduction to
General Relativity, }2$^{nd}$ Edition, McGraw-Hill, New York.

Aguirregabiria, J.M. et al. (1996) - GRG \textbf{28}, 1393.

Albrow, M.G.(1973) - Nature, \textbf{241},56.

Albrecht, A.; Magueijo, J.(1998) \textit{A time varying speed of light as a
solution to cosmological puzzles}-preprint.

\bigskip\ Anderson,J.D. et al.(2002)-\textit{ "Study of the anomalous
acceleration of Pioneer 10 and 11".-} Phys. Rev. D 65, 082004 (2002)

Arbab, A. I. (2004) - \textit{Quantum Universe and the Solution to the
Cosmological Problems,} GRG \textbf{36} , 3565.

\bigskip Bahcall, J.N. ; Schmidt, M. (1967) - Physical Review Letters
\textbf{19}, 1294.

Banerjee, N.; Sen, S. (1997) - Pramana J.Phys., \textbf{49}, 609.

Barrow, J.D. (1988) - \textit{The Inflationary Universe}, in
\textit{Interactions and Structures in Nuclei, }pp. 135-150,\ ed. by R. Blin
Stoyle and W.D. Hamilton, Adam Hilger, Bristol.

Barrow, J.D. (1990) - in \textit{Modern Cosmology in Retrospect}, ed. by B.
Bertotti, R.Balbinot, S.Bergia and A.Messina. CUP, Cambridge.

Barrow, J.D. (1997) - \ \ \textit{Varying G and Other Constants}\ , Los Alamos
Archives http://arxiv.org/abs/gr-qc/9711084 v1 27/nov/1997.

Barrow, J.D. (1998) - in \textit{Particle Cosmology, }Proceedings RESCEU
Symposium on Particle Cosmology, Tokyo, Nov 10-13, 1997, ed. K. Sato, T.
Yanagida, and T. Shiromizu, Universal Academic Press, Tokyo, pp. 221-236.

Barrow, J.D. (1998a) - Cosmologies with Varying Light Speed, Los Alamos
Archives http://arxiv.org/abs/Astro-Ph/9811022 v1 .

Barrow, J.D. ; Magueijo, J. (1999) - Solving the Flatness and Quasi-Flatness
Problems in Brans-Dicke Cosmologies with varying Light Speed. Class. Q.
Gravity, 16, 1435-54.

Bekenstein, J.D. (1982) - Physical Review \textbf{D25}, 1527.

Berman, M. S. (1981, unpublished) - M.Sc. thesis, Instituto Tecnol\'{o}gico de
Aeron\'{a}utica, S\~{a}o Jos\'{e} dos Campos, Brazil.Available online, through
the federal government site \ www.sophia.bibl.ita.br/biblioteca/index.html
(supply author%
\'{}%
s surname and keyword may be "pseudotensor"or "Einstein").

Berman,M.S. (1983)-\textit{Special Law of Variation for Hubble%
\'{}%
s Parameter,}Nuovo Cimento \textbf{74B,}182-186.

Berman, M.S. (1992) - \textit{Large Number Hypothesis} - International Journal
of Theoretical Physics, \textbf{31}, 1447.

Berman, M.S. (1992a) - \textit{A Generalized Large Number Hypothesis} -
International Journal of Theoretical Physics, \textbf{31}, 1217-19.

\bigskip Berman, M. S. (1994) - \textit{A Generalized Large Number Hypothesis}
- Astrophys. Space Science, \textbf{215}, 135-136.

Berman, M.S. (1996) - \textit{Superinflation in G.R. and B.D. Theories: An
eternal Universe} - International Journal of Theoretical Physics \textbf{35}, 1789.

\bigskip Berman, M. S. (2006) - \textit{Energy of Black-Holes and Hawking%
\'{}%
s Universe}, in Chapter 5 of \textit{Trends in Black Hole Research}, ed by
Paul V. Kreitler, Nova Science, New York.

Berman, M. S. (2006a) - \textit{Energy, Brief History of Black-Holes,and
Hawking%
\'{}%
s Universe}, in Chapter 5 of: \textit{New Developments in Black Hole
Research}, ed by Paul V. Kreitler, Nova Science, New York.

\bigskip Berman, M. S. (2007) - \textit{Introduction to General Relativity and
the Cosmological Constant Problem, }Nova Science, New York.

Berman, M. S. (2007a) - \textit{Introduction to General Relativistic and
Scalar Tensor}

\textit{Cosmologies }, Nova Science, New York.

Berman, M.S. (2007b) - \textit{The Pioneer Anomaly and a Machian Universe} -
Astrophysics and Space Science, \textbf{312}, 275. Los Alamos Archives, http://arxiv.org/abs/physics/0606117.

Berman, M. S. (2007c) - \textit{Gravitomagnetism and Angular Momenta of
Black-Holes} - RevMexAA, \textbf{43,} 297-301.

Berman, M. S. (2007d) - \textit{Is the Universe a White-hole?} - Astrophysics
and Space Science, \textbf{311}, 359.

Berman, M. S. (2008a) - \textit{A General Relativistic Rotating Evolutionary
Universe, }Astrophysics and Space Science, \textbf{314, }319-321.

Berman, M. S. (2008b) - \textit{A General Relativistic Rotating Evolutionary
Universe - Part II, }Astrophysics and Space Science, \textbf{315, }367-369.

\bigskip Berman, M.S. (2008c) - \textit{A Primer in Black Holes, Mach's
Principle and Gravitational Energy, }Nova Science, New York.

Berman, M.S. (2008d) - \textit{On the Machian Origin of Inertia}, Astrophysics
Space Science, \textbf{318}, 269-272. Los Alamos Archives, http://arxiv.org/abs/physics/0609026

Berman, M.S. (2008e) - \textit{General Relativistic Machian Universe,
}Astrophysics and Space Science, \textbf{318}, 273-277.

Berman, M.S. (2008f) - \textit{Shear and Vorticity in a Combined
Einstein-Cartan-Brans-Dicke Inflationary Lambda Universe, }Astrophysics and
Space Science, \textbf{314, }79-82. For a preliminary report, see Los Alamos
Archives, http://arxiv.org/abs/physics/0607005

\bigskip Berman, M.S. (2009g) - \textit{Gravitons, Dark Matter}, \textit{and
Classical Gravitation,}AIP Conference Proceedings \textbf{1168,}%
1068-1071\textit{.}Preliminary version,see\textit{ }Los Alamos Archives
http://arxiv.org/abs/0806.1766 .

Berman, M.S. (2009) - \textit{General Relativistic Singularity-free
Cosmological Model, }Astrophysics and Space Science, \textbf{321}, 157-160.

\bigskip Berman, M.S.(2009a) - \textit{Simple Model with time-varying
fine-structure "constant", RevMexAA }\textbf{45, }139-142.

Berman, M.S. (2009b) - \textit{Entropy of the Universe, }International Journal
of Theoretical Physics, \textbf{48}, 1933. DOI 10.1007/s10773-009-9966-4 . Los
Alamos Archives http://arxiv.org/abs/0904.3135.

Berman, M.S. (2009c) - \textit{On the zero-energy Universe, }International
Journal of Theoretical Physics, \textbf{48}, 3278.

Berman, M.S. (2009d) - \textit{Why the initial infinite singularity of the
Universe is not there?, }International Journal of Theoretical Physics,
\textbf{48}, 2253.

Berman, M.S. (2009e) - \textit{On Sciama's Machian Cosmology, }International
Journal of Theoretical Physics, \textbf{48}, 3257.

Berman, M.S. (2009f) - \textit{General Relativistic Singularity-Free
Cosmological Model, }Astrophysics and Space Science, \textbf{321}, 157. Los
Alamos Archives http://arxiv.org/abs/0904.3141

\bigskip Berman, M.S. (2010) - \textit{Simple Model with time-varying
fine-structure "constant" - Part II, RevMexAA }\textbf{46, }23-28.

Berman, M.S.; Gomide, F.M.(1988) - \textit{Cosmological Models with Constant
Deceleration Parameter - }GRG \ \textbf{20,}191-198.

\bigskip Berman,M.S.;Gomide,F.M.(2010)-old version of the present submitted
paper to this Journal.See Los Alamos Archives,arxiv:1011.4627v3

Berman,M.S.;Gomide,F.M.(2011)-\textit{On the Rotation of the Zero-Energy
Expanding Universe}, in "The Big-Bang:Theory, Assumptions and Problems", ed.
by O`Connell and Hale,Nova Science,N.Y., to be published.

Berman,M.S.;Gomide,F.M.(2011a)-See Los Alamos Archives,arxiv:1106.5388v3

Berman, M.S.; Trevisan, L.A. (2001)\ - Los Alamos Archives http://arxiv.org/abs/gr-qc/0112011

\begin{description}
\item Berman, M.S.; Trevisan, L.A. (2001a) - Los Alamos Archives http://arxiv.org/abs/gr-qc/0111102

\item Berman, M.S.; Trevisan, L.A. (2001b) - Los Alamos Archives http://arxiv.org/abs/gr-qc/0111101

\item Berman, M.S.; Trevisan, L.A. (2010) - International Journal of Modern
Physics, \textbf{D19, }1309-1313. For a preliminary version, see Los Alamos
Archives http://arxiv.org/abs/gr-qc/0104060
\end{description}

Berry, M.V. (1989) - \textit{Principles of Cosmology and Gravitation, }Adam
Hilger, Bristol.

Birch, P. (1982) - Nature, \textbf{298}, 451.

Birch, P. (1983) - Nature, \textbf{301}, 736.

\bigskip Carmeli,M. ;Leibowitz,E. ;Nissani, N.
(1990)-\textit{Gravitation:SL(2,C) Gauge Theory and Conservation Laws,}World
Scientific, Singapore.

Cooperstock, F.I. (1994) - GRG \textbf{26}, 323.

Cooperstock, F.I.; Israelit,M. (1995) - \textit{Foundations of Physics},
\textbf{25}, 631.

\bigskip Cooperstock, F.I.;Faraoni,V.(2003) - Ap.J. 587,483.

\bigskip de Sabbata, V.; Gasperini, M. (1979) - \textit{Lettere al Nuovo
Cimento, }\textbf{25, }489.

\begin{description}
\item de Sabbata, V.; Sivaram, C. (1994) - \textit{Spin and Torsion in
Gravitation, }World Scientific, Singapore.

\item Dicke, R.H. (1964) - \textit{The many faces of Mach, }in
\ \textit{Gravitation and Relativity,}ed. by Chiu,H.-Y. and
Hoffmann,W.F.\textit{, } Benjamin , New York.

\item Dicke, R.H. (1964a) - \textit{The significance for the solar system of
time-varying gravitation, }in \ \textit{Gravitation and Relativity,}ed. by
Chiu,H.-Y. and Hoffmann, W.F., Benjamin , New York.

\item Dirac, P.A.M. (1938) - \textit{Proceedings of the Royal Society
}\textbf{165} A, 199.

\item Dirac, P.A.M. (1974) - \textit{Proceedings of the Royal Society
}\textbf{A338,} 439.
\end{description}

\bigskip\bigskip Eddington, A.S. (1933) - \textit{Expanding Universe, }CUP, Cambridge.

\bigskip Eddington, A.S. (1935) - \textit{New Pathways in Science}. Cambridge
University Press, Cambridge.

Eddington, A.S. (1939) - Sci. Progress, London, \textbf{34}, 225.

Feng, S.; Duan, Y. (1996) - Chin. Phys. Letters, \textbf{13}, 409.

Feynman, R. P. (1962-3) - \textit{Lectures on Gravitation} , Addison-Wesley, Reading.

Freud, P.H.(1939) - Ann. Math, \textbf{40}, 417.

Garecki, J.\ (1995) - GRG, \textbf{27}, 55.

\bigskip Godlowski,W.;et al.(2004)-Los Alamos Archives astro-ph/0404329.See
also GERG Journal.

Godlowski,W.(2011)- \ Los Alamos Archives, arxiv:1103.5786v3

Gomide, F.M. (1976) - \textit{Lett. Nuovo Cimento} \textbf{15}, 515.

Gomide, F.M.; Berman, M.S.; Garcia, R.L. (1986) - RevMexAA, \textbf{12}, 46.

Gomide, F.M.; Uehara, M. (1981) - Astronomy and Astrophysics, \textbf{95}, 362.

Guth, A. (1981) - Phys. Rev. \textbf{D23}, 347 .

Guth, A. (1998) - \textit{The Inflationary Universe,}\ Vintage, New York, page 12.

Halliday, D.; Resnik, R.; Walker, J. (2008) - \textit{Fundamentals of Physics,
\ }Wiley. 8$^{th}$ Edition. New York.\textit{ }

\bigskip Hawking, S. (1996) - \textit{The Illustrated A Brief History of Time,
}Bantam Books, New York, pages 166-167.

Hawking, S. (2001) - \textit{The Universe in a Nutshell, }Bantam Books, New
York, pages 90-91.

Hawking, S. (2003) - \textit{The Illustrated Theory of Everything, }Phoenix
Books, Beverly Hills, page 74.

\bigskip Iorio,L.(2010)-JCAP 08,030.

Islam, J.N. (1985) - \textit{Rotating Fields in General Relativity, }CUP, Cambridge.

Johri, V.B.; et al. (1995) - GRG, \textbf{27}, 313.

Katz, J. (1985) - Classical and Quantum Gravity \textbf{2}, 423.

Katz, J. (2006) - Private communication.

Katz, J.; Ori, A. (1990) - Classical and Quantum Gravity\textbf{ 7}, 787.

Katz, J.; Bicak, J.; Lynden-Bell, D. (1997) - Physical Review \textbf{D55}, 5957.

Landau, L.; Lifshitz, E. (1975) - \textit{The Classical Theory of Fields},
4th. Revised ed.; Pergamon, Oxford.

Moffat, J.W. (1993) - International Journal of Modern Physics, \textbf{D2}, 351.

\bigskip Ni,Wei-Tou(2008)-Progress Theor. Phys.Suppl.\textbf{172,}49-60.

Ni,Wei-Tou (2009)-Int.Journ.Modern Phys.\textbf{A24,}3493-3500.

Papapetrou, A. (1974) - \textit{Lectures on General Relativity}, Reidel, Boston.

Radinschi, I. (1999) - Acta Phys. Slov., \textbf{49}, 789. Los Alamos
Archives, gr-qc/0008034.

Reitz, J.R.; Milford, F.J.; Christy, R.W. (1979) - \textit{Foundations of
Electromagnetic Theory}, Addison-Wesley. Reading.

Rosen, N.(1994) - \textit{Gen. Rel. and Grav.} \textbf{26}, 319.

Rosen, N.(1995) - GRG, \textbf{27}, 313.

Sabbata, de V.; Gasperini, M. (1979) - \textit{Lettere al Nuovo Cimento,
}\textbf{25}, 489.

Schmidt, B.P. (1998) \ - Ap.J., \textbf{507}, 46-63.

Schwarzschild, B. (2001)- Physics Today, \textbf{54} (7), 16.

Sciama, D.W. (1953) - M.N.R.A.S., \textbf{113}, 34.

So, L.L.; Vargas, T. (2006) - Los Alamos Archives, gr-qc/0611012 .

Tryon, E.P.(1973) - Nature, \textbf{246}, 396.

\begin{description}
\item Webb, J.K; et al. (1999) --Phys. Rev. Lett. \textbf{82}, 884.

\item Webb, J.K; et al. (2001) -- Phys Rev. Lett. \textbf{87,} 091301.
\end{description}

Weinberg, S.\ (1972) - \textit{Gravitation and Cosmology, }Wiley. New
York.\bigskip

Will, C. (1987) - in \textit{300 Years of Gravitation}, ed. by W.Israel and
S.Hawking, CUP, Cambridge.

Will, C. (1995) - in \textit{General Relativity - Proceedings of the 46}%
$^{th}$ \textit{Scotish Universities Summer School in Physics, }ed. by G.Hall
and J.R.Pulhan, IOP/SUSSP Bristol.

\bigskip Xulu, S. (2000) - International Jounal of Theoretical Physics,
\textbf{39}, 1153. Los Alamos Archives, gr-qc/9910015 .

York Jr, J.W. (1980) - \textit{Energy and Momentum of the Gravitational
Field}, in \textit{A Festschrift for Abraham Taub}, ed. by F.J. Tipler,
Academic Press, N.Y.\bigskip

\end{document}